\documentclass[aps,prl,reprint,twoside,tightenlines,showpacs,amsmath,amssymb,nofootinbib,floatfix,superscriptaddress]{revtex4-1}
\usepackage{graphicx}
\usepackage{amsfonts}
\usepackage{color}
\usepackage{multibib}
\usepackage[colorlinks=true,urlcolor=blue,linkcolor=blue,citecolor=blue]{hyperref}
\usepackage[T1]{fontenc}
\usepackage{lmodern}

\begin{document}

\title{\boldmath Signature of an $h_1$ state in the $J/\psi \to \eta h_1 \to \eta K^{*0}\bar{K}^{*0}$ decay}
\date{\today}
\author{Ju-Jun Xie}
\affiliation{Institute of Modern Physics, Chinese Academy of
Sciences, Lanzhou 730000, China} 
\affiliation{State Key Laboratory of Theoretical Physics,
Institute of Theoretical Physics, \\ Chinese Academy of Sciences,
Beijing 100190, China}
\affiliation{Departamento de
F\'{\i}sica Te\'orica and IFIC, Centro Mixto Universidad de
Valencia-CSIC Institutos de Investigaci\'on de Paterna, Aptdo.
22085, 46071 Valencia, Spain }
\author{M.~Albaladejo}
\author{E.~Oset}
\affiliation{Departamento de
F\'{\i}sica Te\'orica and IFIC, Centro Mixto Universidad de
Valencia-CSIC Institutos de Investigaci\'on de Paterna, Aptdo.
22085, 46071 Valencia, Spain }

\begin{abstract}
The BES data on the $J/\psi \to \eta K^{*0}\bar{K}^{*0}$ reaction show a clear enhancement in the $K^{*0}\bar{K}^{*0}$ mass distribution close to the threshold of this channel. Such an enhancement is usually a signature of a $L=0$ resonance around threshold, which in this case would correspond to an $h_1$ state with quantum numbers $I^G(J^{PC})=0^-(1^{+-})$. A state around $1800\ \text{MeV}$ results from the interaction of the $K^* \bar{K}^*$ using the local hidden gauge approach. We show that the peak observed in $J/\psi \to \eta K^{*0}\bar{K}^{*0}$ naturally comes from the creation of this $h_1$ state with mass and width around $1830\ \text{MeV}$ and $110\ \text{MeV}$, respectively. A second analysis, model independent, corroborates the first result, confirming the relationship of the enhancement in the invariant mass spectrum with the $h_1$ resonance.
\end{abstract}

\maketitle

The decay $J/\psi \to \eta K^{*0}\bar{K}^{*0}$ was measured for
the first time by the BES Collaboration~\cite{BESdata} with the aim of
searching for the $Y(2175)$ resonance through the decay $J/\psi \to \eta
Y(2175) \to \eta K^{*0}\bar{K}^{*0}$. However, no clear enhancement in the $K^{*0}\bar{K}^{*0}$ mass distribution was found near $2.175$ GeV.  Although unknown at the time of the experiment, it is easier to understand at present that the $Y(2175)$, now catalogued as $\phi(2170)$ in the PDG~\cite{pdg2012}, with quantum numbers $I^G(J^{PC})=0^-(1^{--})$, does not couple to $K^{*0}\bar{K}^{*0}$. The resonance couples strongly to the $\phi f_0(980)$ \cite{Aubert:2006bu}, and it was found in Refs.~\cite{MartinezTorres:2008gy, AlvarezRuso:2009xn, Coito:2009na}, using Fadeev calculations and related methods, that it was made of a cluster of $\phi K \bar{K}$, with the $K\bar{K}$ highly correlated into an $f_0(980)$. On the other hand, since the $J/\psi$ and the $\eta$ mesons have quantum numbers $0^-(1^{--})$ and $0^+(0^{-+})$, respectively, the decay $J/\psi \to \eta K^{*0}\bar{K}^{*0}$ constitutes the ideal reaction to look for an $h_1$ state, with quantum numbers $0^-(1^{+-})$, coupling to an $s$--wave $K^* \bar{K^*}$ pair.

Both on the theoretical and experimental sides, the study of $h_1$
states above $1.5$ GeV is very scarce~\cite{pdg2012}. In Ref.~\cite{gengoset}, the work of Ref.~\cite{Molina:2008jw} on the $\rho\rho$ interaction was extended to $\text{SU}(3)$ using the local hidden gauge formalism for vector--vector interaction and a unitary approach in coupled channels. This interaction generates resonances in different strangeness-isospin-spin channels. In the $0^-(1^{+-})$ sector, a resonance was found with mass and width around $1800$ and $80$ MeV, respectively. This resonance is dynamically generated from the
interaction of $K^*$ and $\bar{K}^*$. As shown in Ref.~\cite{gengoset}, given its quantum numbers, it cannot couple to other vector--vector or pseudoscalar--pseudoscalar channels, which makes its
observation difficult. This state cannot be clearly identified with any of the $h_1$ states listed in the PDG, and it is not found in
Ref.~\cite{GarciaRecio:2010ki} (or, rather, it is pushed down to lower energy).  Reference~\cite{GarciaRecio:2010ki} uses different dynamics, based on spin-flavour $\text{SU(6)}$ symmetry, and a regularization of the loops which is different from the one in Ref.~\cite{Molina:2008jw}. Both approaches share qualitatively many features, but differ somewhat in the numerical results. A different dynamical approach to the vector-vector interactions is provided in \cite{Albaladejo:2008qa}, using a massive Yang-Mills field theoretical framework. On the experimental side, in an analysis of $p\bar{p}$ annihilation channels done in Ref.~\cite{Anisovich:2011sva}, an $h_1$ state with mass and width $1965$ and $345$ MeV is found, very far from the one predicted in Ref.~\cite{gengoset}.

On the other hand, a careful inspection of the BES results in Ref.~\cite{BESdata},
shows an enhancement in the invariant mass spectrum of the
$K^{*0}\bar{K}^{*0}$ around $1850$ MeV (see more details from Fig.~4 of Ref.~\cite{BESdata}), which might hint to a sizeable
contribution from an $h_1$ state strongly coupling to the
$K^{*0}\bar{K}^{*0}$ channel. However, as explained above,
the search of the $\phi(2170)$ resonance was the main aim of
Ref.~\cite{BESdata}, and no attention was paid to this enhancement.

\begin{figure}[t!]\centering
\includegraphics[width=0.47\textwidth]{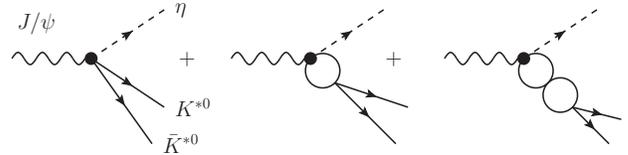}
\caption{Diagrammatic representation of the $J/\psi \to \eta
K^{*0}\bar{K}^{*0}$ decay. \label{feydgm}}
\end{figure}

In the present work, following the formalism of Ref.~\cite{gengoset},
we shall make a first study of the role of the $h_1$ state (which is
dynamically generated by the $K^*$ and $\bar{K^*}$ interaction) in
the $J/\psi \to \eta K^{*0}\bar{K}^{*0}$ decay. The decay mechanism
is shown in Fig.~\ref{feydgm}. Let us denote by $V_P$ the bare production vertex
for $J/\psi \to \eta K^{*0}\bar{K}^{*0}$. We assume that this bare vertex is of a short range nature, {\it i.e.}, just a coupling constant in the field theory language. To take into account the final state interaction of the $K^* \bar{K}^*$ pair (which, in the end, will generate the $h_1$ resonance), one has to make the resummation of the diagrams shown in Fig.~\ref{feydgm}. Let us also denote by $t(M_\text{inv}^2)$ and $v(M_\text{inv}^2)$ (where $M_\text{inv}$ is the $K^* \bar{K}^*$ invariant mass) the full amplitude and the bare interaction vertex, respectively, of an $I=0$ $K^* \bar{K}^*$ pair. These are related by:
\begin{equation}\label{eq:tmat}
t = v+v\widetilde{G}t = v ( 1+ \widetilde{G} t) =(1-v\widetilde{G})^{-1} v = (v^{-1} - \widetilde{G})^{-1}~,
\end{equation}
where $\widetilde{G}(M_\text{inv}^2)$ is the loop function for the $K^* \bar{K}^*$ pair. This function is divergent, and it can be regularized both with a cutoff prescription or with
dimensional regularization in terms of a subtraction
constant~\cite{Oller:2000fj}. Here we shall make use of the
dimensional regularization scheme, which introduces a subtraction constant, $a(\mu)$, where $\mu$ is a regularization scale, taken here as $\mu = 1\ \text{GeV}$ (notice that there is only one free parameter). Moreover, since the $K^*$ and
$\bar{K^*}$ have large total decay widths, they should be taken into
account. For that purpose, the $\widetilde{G}$ function is the loop function for two stable particles of masses $m_1$ and $m_2$, $G(M_\text{inv}^2,m_1^2,m_2^2)$, but convoluted in the masses $m_1$ and $m_2$ with the mass distribution of the two vectors, as done in Refs.~\cite{gengoset, Molina:2008jw}. The distribution is considered in the range $m_{1,2} = m_{K^*} \pm 2 \Gamma_{K^*}$, where $m_K^*$ and $\Gamma_{K^*}$ are the nominal mass and width of the $K^*$ meson, respectively. Explicit expressions for $v$ and $\widetilde{G}$ can be found in Ref.~\cite{gengoset}.

With all these definitions, the full amplitude $t_P(M_\text{inv}^2)$ for the process $J/\psi \to \eta K^* \bar{K}^*$ can be written, according to the diagrams in Fig.~\ref{feydgm}, as:
\begin{equation}
t_P = V_P \left(1 + \widetilde{G}(M_\text{inv}^2) t(M_\text{inv}^2) \right) = V_P \frac{t(M_\text{inv}^2)}{v(M_\text{inv}^2)}~,
\end{equation}
where the last equality follows from Eq.~\eqref{eq:tmat}. Then we can easily get the $K^{*0} \bar{K}^{*0}$ invariant mass spectrum for the $J/\psi \to \eta K^{*0} \bar{K}^{*0}$ as \cite{Nacher:1998mi}:
\begin{eqnarray}
\frac{d\Gamma}{d M_{\text{inv}}} &=& \frac{C}{\left\lvert v(M_\text{inv}^2) \right\rvert^2} \frac{p_1
\widetilde{p}_2}{M_{J/\psi}} \left\lvert t(M_\text{inv}^2) \right\rvert^2, \label{eq:dgdm}
\end{eqnarray}
with $C$ a global constant factor (that absorbs the unknown vertex $V_P$) which will be fitted to the data, and $p_1$ is the $\eta$ momentum in the $J/\psi$ rest
frame,
\begin{equation}
p_1 = \frac{\lambda^{1/2}(M^2_{J/\psi}, m^2_{\eta},
M^2_\text{inv})}{2M_{J/\psi}},
\end{equation}
where $\lambda(x,y,z)$ is the K\"ahlen or triangle function. In Eq.~\eqref{eq:dgdm}, $\widetilde{p}_2$ is the momentum of the $K^{*0}$ in the $K^{*0} \bar{K}^{*0}$ center of mass system, but, as for the $\widetilde{G}$ function, we must take into account the large width of the $K^*$ meson. That is, $\widetilde{p}_2$ is the convolution with the mass distribution of the two $K^{*0}$ (in the range $m_{1,2} = m_{K^*} \pm 2\Gamma_{K^*}$), of the momentum $p_2$, given by:
\begin{equation}
p_2 = \frac{\lambda^{1/2}(M^2_\text{inv},m^2_1,m^2_2)}{2M_\text{inv}}.
\end{equation}

The $K^{*0}\bar{K}^{*0}$ invariant mass spectrum, Eq.~\eqref{eq:dgdm}, depends on the amplitude $t$, and the most important ingredient in this amplitude is the $K^*\bar{K^*} \to K^*\bar{K^*}$ transition potential $v$. We will discuss two approaches to this potential in this work. In principle, this potential is fixed from the hidden gauge unitary approach of Ref.~\cite{gengoset}. There, it was found that:
\begin{equation}\label{eq:potential}
v = \left( 9 + b \left( 1-\frac{3M^2_\text{inv}}{4 m^2_{K^*}} \right) \right)g^2~,
\end{equation}
where $g = m_\rho / 2f$, being $m_\rho$ and $f$ the mass of the $\rho$ meson and the pion weak decay constant, respectively. The term $9g^2$ comes from the four vector contact term, whereas the term proportional to $b$ comes from the exchange of vector mesons. The constant $b$ is determined by the masses of the vector mesons ($\rho$, $\omega$, $\phi$ and $K^*$) and its value turns out to be $b=6.8$. If one uses Eq.~\eqref{eq:potential} there are two free parameters for calculating the $K^{*0}\bar{K}^{*0}$ invariant mass spectrum of Eq.~\eqref{eq:dgdm}. One is the global constant $C$, which does not influence the amplitude $t$. The other one is the subtraction constant $a(\mu)$, that completely determines now the amplitude $t$. This constant, in the absence of data and by similarity with other channels, was fixed in Ref.~\cite{gengoset} to $a(\mu) = -1.7$, but this value does not allow to obtain a good reproduction of the BES data for the $K^{*0}\bar{K}^{*0}$ invariant mass spectrum. It produces a peak that is narrower and lower in mass than the bump seen in the data. Instead, we find a good agreement with the data using the rather natural range $-1.0 < a(\mu) < -0.6$. The second possibility is to consider a constant potential (in contrast to the one in Eq.~\eqref{eq:potential}, which depends on $M_\text{inv}^2$), fitted to reproduce the data. A constant potential is acceptable here given the small range of $M_\text{inv}$ considered in this work ($1.7\ \text{GeV} < M_\text{inv} < 2.1\ \text{GeV}$), and would render our analysis more model independent. By considering a constant potential, the subtraction constant $a(\mu)$ is no longer a free parameter. This is so because it is an additive term in the $\widetilde{G}$ function, and then any shift in this subtraction constant can be exactly absorbed in the constant potential, as can be seen in Eq.~\eqref{eq:tmat}. For this reason, we fix $a(\mu) = -0.8$ in this case (the central value of the range used above). Then, when using the  constant potential, two free parameters need to be determined: the global constant $C$, and the potential $v$ itself. 

\begin{figure}[t!]\centering
\includegraphics[width=0.49\textwidth]{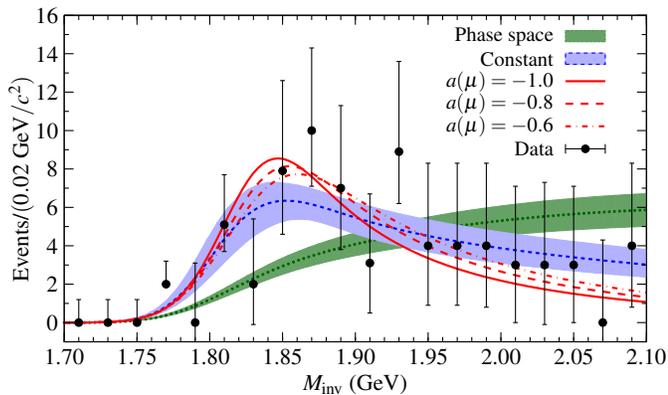}
\caption{(Color online) The $K^{*0}\bar{K}^{*0}$ invariant mass
spectrum of $J/\psi \to \eta K^{*0}\bar{K}^{*0}$ decay. The data points are taken from
Ref.~\cite{BESdata}. The different lines represent the output of Eq.~\eqref{eq:dgdm} for the different approaches considered in this work. The short-dashed line and the associated error band (light blue) represent the results of the constant potential. The (red) solid, long-dashed and dot-dashed lines represent the results for the potential in Eq.~\eqref{eq:potential} with $a(\mu)=-1.0$, $-0.8$ and $-0.6$, respectively. Finally, the (green) dotted line, and the associated error band (dark green) is the prediction for phase space alone.\label{fig:dgdm}}
\end{figure}

In both approaches, the free parameters are fitted by means of a $\chi^2$ function to reproduce the $K^{*0}\bar{K}^{*0}$ invariant mass spectrum data of the BES Collaboration, shown in Fig.~\ref{fig:dgdm}. Actually, the points in the latter represent the number of events once the sideband events are subtracted from the experimental points for the $K^{*0}K^{*0}$ mass distribution given in Ref.~\cite{BESdata}.\footnote{The sideband events are removed in Ref.~\cite{BESdata} to evaluate the total rate of the process $J/\psi \to \eta K^{*0}K^{*0}$, but the data for the mass distribution shown in Fig.~4 of Ref.~\cite{BESdata} are given without this subtraction.} As a further constraint, we consider in the $\chi^2$ function the total number of events in the invariant mass range considered. 

\begin{table}[t!]\centering
\caption{Values of some of the parameters used or determined in this
work.\label{tab:params}}
\begin{ruledtabular}\begin{tabular}{ccccc}
Potential & $C$ (GeV$^{-1}$) & $a_{\mu}$ & $v/g^2$ & $\chi^2/\text{d.o.f.}$ \\
\hline
Constant     & $42 \pm 6$ & $-0.8$ & $-6.2 \pm 1.2$           & $0.45$ \\
Hidden gauge & $42 \pm 6$ & $-1.0$ & Eq.~\eqref{eq:potential} & $0.56$ \\
Hidden gauge & $53 \pm 7$ & $-0.8$ & Eq.~\eqref{eq:potential} & $0.47$ \\
Hidden gauge & $67 \pm 9$ & $-0.6$ & Eq.~\eqref{eq:potential} & $0.42$ \\
\end{tabular}\end{ruledtabular}
\end{table}
We start with the discussion of the second approach (constant potential). The best fit values of the potential $v$ and the global constant $C$ are shown in Table \ref{tab:params}. The blue short-dashed line in Fig.~\ref{fig:dgdm} represents the spectrum obtained with these parameters. The errors of the parameters as well as the error band of the curve are given by the condition $\chi^2 \leqslant \chi^2_\text{min} + 1$, where $\chi^2_\text{min}$ is the minimum $\chi^2$. It is worth noting, by inspection of Fig.~\ref{fig:dgdm}, that the reproduction of the data is good. The $\chi^2/\text{d.o.f.}$ is rather small, given the large errors in the experimental data. We stress that this approach is rather model independent, since we do not assume any underlying model for the transition potential $v$, but take it to be a constant. We now consider the approach in which the potential is taken from the local hidden gauge theory, Eq.~\eqref{eq:potential}, for three different values of the subtraction constant, $a(\mu)=-1.0$, $-0.8$ and $-0.6$. The results are shown in Fig.~\ref{fig:dgdm} with solid, long-dashed, and dot-dashed lines, respectively. The fitted global constant for each case is shown in Table~\ref{tab:params}. The $\chi^2/\text{d.o.f.}$ is similar to that obtained with the constant potential and, inspecting Fig.~\ref{fig:dgdm}, it is also clear that the reproduction of the data is also good.

Both approaches (constant and local hidden gauge potentials) give results in good agreement with the data, and give hints for the presence of a resonance. Yet, we would like to see if the experimental $K^{*0}\bar{K}^{*0}$ invariant mass spectrum in Fig.~\ref{fig:dgdm} can be accounted for by solely phase space. This is easily achieved by putting $t=v=1$ in Eq.~\eqref{eq:dgdm}, and adjusting $C$. This turns out to be $C=48 \pm 7\ \text{GeV}^{-1}$, and the fit has a larger $\chi^2/\text{d.o.f} = 0.9$. The dotted curve and the associated error band (dark green band) in Fig.~\ref{fig:dgdm} represent this result. It is easy to see that the inclusion of an $h_1$ state, generated dynamically from the $K^* \bar{K^*}$ interaction, is crucial to achieve a fairly good description of the BES data. The phase space alone clearly does not follow the trend of the data, even the corresponding
$\chi^2/\text{d.o.f} = 0.9$ could be statistically acceptable, which is due to the large errors of the experimental data. But one observes that below $1.93\ \text{GeV}$ almost every datum is above the phase space band, while they are mostly below for energies above that one.

\begin{figure}[t!]\centering
\includegraphics[width=0.49\textwidth]{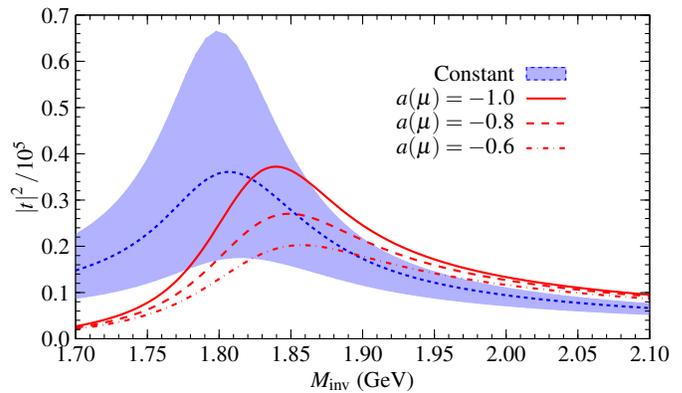}
\caption{(Color online) The modulus squared of the amplitude, $\left\lvert t \right\rvert^2$, for
$K^*\bar{K^*} \to K^*\bar{K^*}$, for the different approaches considered in this work. The notation of the lines as in Fig.~\ref{fig:dgdm}.\label{fig:ts}}
\end{figure}

Now, we use the parameters obtained in the fits to see if the amplitude $t$ obtained by means of Eq.~\eqref{eq:tmat} with both approaches has or has not a resonance. For this, we plot the modulus squared of the scattering
amplitude of $K^*\bar{K^*} \to K^*\bar{K^*}$ in
Fig.~\ref{fig:ts}. For the case in which the potential is taken as a constant, the peak is located roughly at $1810\ \text{MeV}$, whereas its width is around $100\ \text{MeV}$. In the local hidden gauge potential case, we find instead $1850\ \text{MeV}$ and $120\ \text{MeV}$. In both cases, the $t$ matrix has a resonant shape, and the differences of about $40\ \text{MeV}$ in the position of the peak can be accepted as  systematic uncertainties of the resonance mass in our analysis. Definitely, an experimental study with more statistics would help reduce these uncertainties. Note that, because of the large error bars in the data, the statistical error band (shown in Fig.~\ref{fig:ts} for the constant potential) is quite broad. However, this barely affects the position of the maximum, which moves by less than $10\ \text{MeV}$.

In summary, we have studied the  $J/\psi \to \eta
K^{*0}\bar{K}^{*0}$ decay with the aim of determining the existence of
an $h_1$ state through $J/\psi \to \eta h_1 \to
\eta K^{*0}\bar{K}^{*0}$ decay. In particular, by using a constant
potential and a local hidden gauge potential for the $K^{*}\bar{K}^* \to
K^{*}\bar{K}^*$ transition, we calculate the distributions of
the differential decay width, with which we obtain the line shape of the
$K^{*0}\bar{K}^{*0}$ invariant mass spectrum. By fitting our results
to the BES experimental data, we find that these can be fairly
described by including the contributions from an $h_1$ state, while
the phase space alone cannot describe the data, especially the
bump structure near threshold. We also show the corresponding
results for the modulus squared of the scattering amplitude of
$K^*\bar{K^*} \to K^*\bar{K^*}$, from where we can get the mass and
width of this $h_1$ state around $M_{h_1} = 1830 \pm  20\ \text{MeV}$ and $\Gamma_{h_1} = 110 \pm 10\ \text{MeV}$, respectively.

The analysis done here shows clearly enough that the BES data, with the enhancement of the $K^{*0}\bar{K}^{*0}$ invariant mass spectrum close to the threshold, call for an $h_1$ resonance with the properties given by our fit. Yet, the data could be considerably improved, and in view of the results of the present paper, it would be most advisable to do so, in order to improve on the present statistics, which would also revert into smaller systematic errors in the determination of the resonance properties.

\section*{Acknowledgments}

This work is partly supported by the Spanish Ministerio de Economia
y Competitividad and European FEDER funds under the contract number
FIS2011-28853-C02-01 and FIS2011-28853-C02-02, and the Generalitat
Valenciana in the program Prometeo, 2009/090. We acknowledge the
support of the European Community-Research Infrastructure
Integrating Activity Study of Strongly Interacting Matter (acronym
HadronPhysics3, Grant Agreement n. 283286) under the Seventh
Framework Programme of EU. This work is also partly supported by the
National Natural Science Foundation of China under grant 11105126.

\bibliographystyle{plain}

\end{document}